\documentclass[aps,prl,twocolumn,superscriptaddress]{revtex4-2}
\usepackage{amsmath,amssymb}
\usepackage{amsfonts}
\usepackage{amsbsy}
\usepackage{wasysym}
\usepackage{marvosym}
\usepackage{graphicx,hyperref,color}
\usepackage{stix}

\newcommand{\Pe}{{\rm Pe}}

\newcommand{\OmA}{\Omega_A}
\newcommand{\OmT}{\omega_T}

\graphicspath{{./figures/}}

\begin{document}

\title{Noisy Pursuit and Pattern Formation of Self-Steering Active Particles}
\author{Segun Goh}
\email{s.goh@fz-juelich.de}
\affiliation{Theoretical Physics of Living Matter, Institute of Biological Information Processing, 
	Forschungszentrum J{\"u}lich, 52425 J{\"u}lich, Germany}
\author{Roland G. Winkler}
\email{r.winkler@fz-juelich.de}
\affiliation{Theoretical Physics of Living Matter, Institute of Biological Information Processing, 
	Forschungszentrum J{\"u}lich, 52425 J{\"u}lich, Germany}
\author{Gerhard Gompper}
\email{g.gompper@fz-juelich.de}
\affiliation{Theoretical Physics of Living Matter, Institute of Biological Information Processing, 
	Forschungszentrum J{\"u}lich, 52425 J{\"u}lich, Germany}
\date{\today}
\begin{abstract}

We consider a moving target and an active pursing agent, modeled as
an intelligent active Brownian particle capable of sensing the instantaneous
target location and adjust its direction of motion accordingly. An analytical and
simulation study in two spatial dimensions reveals that pursuit performance depends on the interplay 
between self-propulsion, active reorientation, and random noise. Noise is found to have two opposing 
effects: (i) it is necessary to disturb regular, quasi-elliptical trajectories around the target, 
and (ii) slows down pursuit by increasing the traveled distance of the pursuer.  We also propose 
a strategy to sort active pursuers according to their motility by circular target trajectories.
\end{abstract}
\maketitle
Motility is an essential source of pattern formation and collective dynamics 
in biological and artificial systems on scales from microbes, cells, and colloids, to microbots and animals \cite{Lauga2009,Vicsek2012,Bechinger2016,Elgeti2015,Palagi2018,Gompper2020,Shaebani2020}. Interacting active 
particles and agents exhibit fascinating collective behaviors, ranging from swarming of microorganisms 
\cite{Kearns2010,Sokolov2012,Wensink2012,Qi2022} to lane-formation of ants \cite{Couzin2003,Ayalon2021} 
and flocking of birds \cite{Cavagna2014,Popkin2016,Vicsek1995,Toner1995}.
Theoretical models of such systems account for three main kinds of interactions in 
addition to self-propulsion, which are steric repulsion, velocity alignment, and hydrodynamics 
\cite{Purcell1977,Lauga2009,Elgeti2015}, where the latter implies a classification into dry and wet active matter.
Simple stochastic models, like the active Brownian particle (ABP) and the run-and-tumble (RTB) model,
have provided a theoretical framework for understanding dry active matter \cite{Cates2015, Bechinger2016}.

Biological microorganisms differ in various important aspects from their active-colloid counterparts,
which is their ability to sense their environment, to process the gathered information, and to steer 
their motion accordingly \cite{Jikeli2015,Harpaz2021}. The adopted response enables biological agents to perform goal-oriented motion, which includes cell motion in wound healing, foraging and prey-searching activities of animals, as well as traffic flows \cite{Gazis1967, Chowdhury2000} and pedestrian dynamics \cite{Seyfried2005,Moussaid2011} in social systems.
Notwithstanding the apparent relevance of the role of information processing in motile systems, 
this has only very recently been started to be taken into account in studies of active matter 
\cite{Baeuerle2018,Lavergne2019,Qian2013,Levis2020,Zhang2021,Alvarez2021}.
Active adaptation should be distinguished from responses of particles in passive systems \cite{Kaspar2021}, 
whose dynamics is governed  by conservative (reciprocal) interactions and external forces.

It is important to note that the process of information gathering can be
non-reciprocal and long ranged, far beyond the range of standard colloidal interactions \cite{Saha2020,Fruchart2021}. 
A minimal cognitive flocking system has been introduced, which assumes that the 
moving entities navigate by using exclusively the instantaneous ``visual information'' they receive 
about the position of other entities \cite{Barberis2016,Bastien2020}.
This includes systems and models, in which particles adjust their propulsion direction toward 
regions of highest particle concentration located inside the vision cone \cite{Barberis2016}. 
Alternatively, particle propulsion can be switched on or off depending on whether the particle number 
within the vision cone is above or below a threshold value \cite{Baeuerle2018, Lavergne2019,Qian2013}. 
Such vision-dependent propulsion mechanisms lead to particle aggregation and swarming without any 
attractive interactions.

\begin{figure}
	\includegraphics[width=8.6cm]{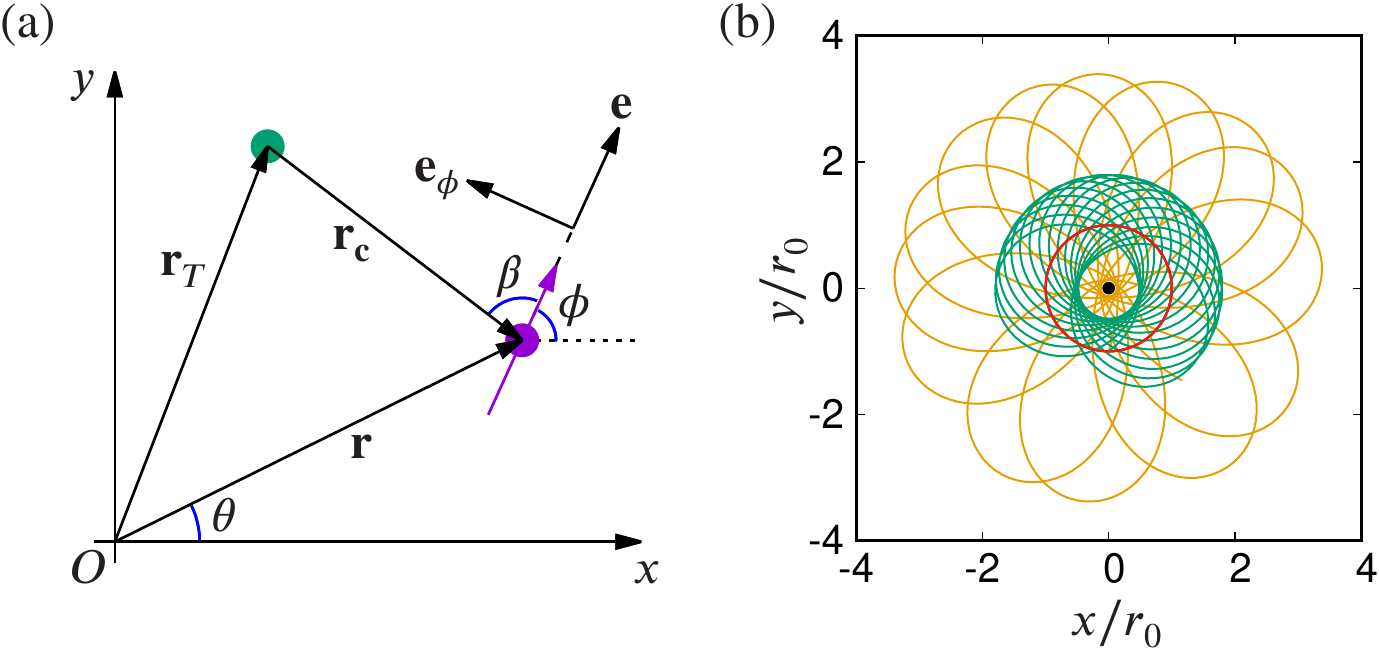}
	\caption{\label{fig:schematics} 
		(a) Schematics of the set-up for a stationary target located at $\mathbf{r}_T$ (green bullet)
			and a pursuer at $\mathbf{r}$ (purple bullet) with the propulsion direction $\mathbf{{e}}$ 
			(purple arrow), and the angle $\beta = \theta - \phi - \pi$.
		(b) Trajectories of pursuers in the absence of noise, with a stationary target at the origin 
		    (black bullet). The red circle represents the marginally stable fixed-point solution 
		    $(r, \beta) = (r_0, \pm\pi/2)$. Two rosette-like motions are shown for different initial 
		    conditions.
		}
\end{figure}

In this study, we consider a minimal model of an intelligent active particle, which is capable of an
active reorientation of its direction of motion toward a moving target. We employ the minimal cognitive
active Brownian particle model introduced in Ref.~\cite{Barberis2016},  assuming perfect sensing, 
such that the pursuer always knows the exact location of the target.  The success of the pursuer to 
approach and reach the target, which moves on a fixed, prescribed trajectory,  depends on the relative 
velocities, the translational and rotational noise of the pursuer, and the strength of the reorientation 
force. In particular, orientational noise plays a fundamental role, as it is, on the one hand, necessary 
to reach a  target,  and, on the other hand, effectively slows down the pursuit by increasing the travel 
distance of the pursuer, and hence requires a larger pursuer velocity, compared to that of the target,
for a successful pursuit. Furthermore, we demonstrate that a target moving on a circular trajectory 
can be used to localize pursuers, and to separate and sort them according to  their velocities.
 
\begin{figure}
	\includegraphics[width=8.6cm]{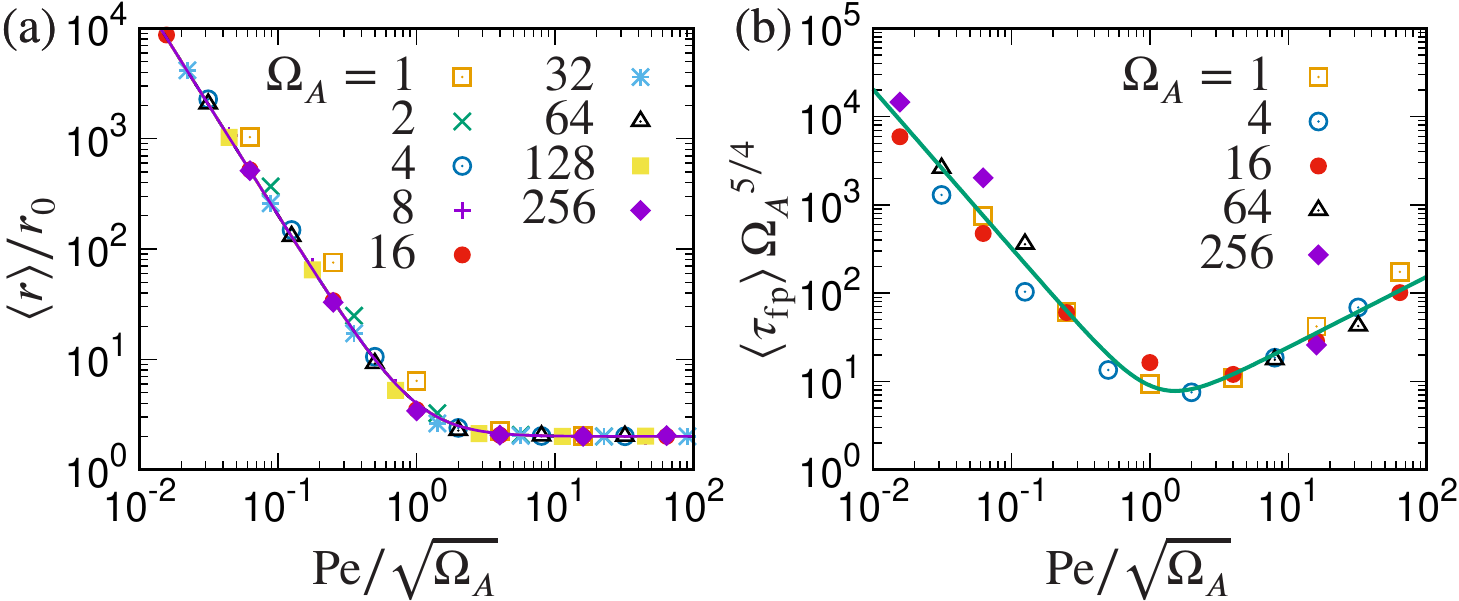}
	\caption{\label{fig:stationary} 
	(a) Mean distance of a pursuer from a stationary target as a function of the scaled P\'eclet number 
	$\Pe/\sqrt{\Omega_A}$ for various $\OmA$. The line represents Eq. \eqref{eq:rave}. 
	(b)  Mean first-passage times  $\langle \tau_{\mathrm{fp}} \rangle$ scaled by $\OmA^{-5/4}$ 
	as a function of $\Pe/ \sqrt{\OmA}$ for various $\OmA$. The solid line interpolates between the power-laws 
	$\langle \tau_{\mathrm{fp}} \rangle \sim \Pe^{-9/5}$ for $\Pe/ \sqrt{\OmA} <1$ and 
	$\langle \tau_{\mathrm{fp}} \rangle \sim \Pe^{4/5}$ for $\Pe/ \sqrt{\OmA} >1$. } 
\end{figure}
 
The translational motion of the intelligent active Brownian pursuer (iABP) is described by the overdamped Langevin equation 
\cite{Fodor2016, Das2018}
\begin{align} \label{eq:position}
\mathbf{\dot{r}} =  \mathbf{v} + \sqrt{2D_T} \ \boldsymbol{\eta}_T,
\end{align}
where $\boldsymbol{\eta}_T$ is a Gaussian white-noise  stochastic process with zero mean and second moment
$\langle \boldsymbol{\eta}_T(t)\cdot \boldsymbol{\eta}_T (t') \rangle = 2\delta (t-t')$, $D_T$ is the 
translation diffusion coefficient, and ${\bf v} = v_0 {\bf e}$ is the active velocity of constant magnitude 
$v_0$ along the propulsion direction $\bf e$ ($|{\bf e}| =1$). The time evolution of  ${\bf e} (t)$ 
is determined by orientational diffusion due to an active process or thermal fluctuations as for a standard
ABP \cite{Cates2015,Bechinger2016}, and an \emph{active adaptation} contribution  $\mathbf{f}_A$, hence, 
\begin{align}
\mathbf{\dot{e}} =   \mathbf{f}_{A} + \sqrt{2D_R}  \boldsymbol{\eta}_R \times {\bf e}, 
\end{align}
with the Gaussian and Markovian stochastic process $\boldsymbol{\eta}_R$ of zero mean, the second moment 
$\langle \boldsymbol{\eta}_R (t) \cdot \boldsymbol{\eta}_R (t') \rangle = 2 \delta (t-t')$, and  
the rotational diffusion coefficient $D_R$.  The active contribution is 
${\bf f}_A = - C_0 \, {\bf e} \times ({\bf e} \times {\bf r_c}/|{\bf r_c}|)$, where ${\bf r}_c = {\bf r} - {\bf r}_T$ is the 
vector  connecting the pursuer at ${\bf r}$ and the target at ${\bf r}_T(t)$. The adaptive force ${\bf f}_A$ 
is orthogonal to $\mathbf{e}$  and ${\bf f}_A =0$ when $\mathbf{e}$ is parallel 
to $\mathbf{r}_c$ \cite{Barberis2016}.

We consider a two-dimensional system and polar coordinates for ${\bf r}$ and ${\bf e}$, as indicated in 
Fig.~\ref{fig:schematics}(a), and introduce dimensionless quantities by measuring length and time in units 
of $r_H =\sqrt{D_T/D_R}$, where $r_H$ is analogous to the Stokes radius of a spherical colloid 
in a fluid, and $D_R$, respectively, which corresponds to $r \to r/r_H$ and $t \to D_R t$. The equations 
of motion then become
\begin{align} 
\dot{r} &= \Pe \cos{(\theta-\phi)}  +\sqrt{2} {\eta}_r, \label{eq:ABP_mov_r} \\
\dot{\theta} &= - \frac{\Pe}{r}\sin{(\theta-\phi)}  +\frac{\sqrt{2}}{r}{\eta}_\theta, \label{eq:ABP_mov_theta}\\
\dot{\phi} &= -\frac{\OmA}{r_c} \left( r \sin{(\theta-\phi)}  - {\bf r}_T \cdot {\bf e}_{\phi} \right)+\sqrt{2} {\eta}_R, \label{eq:ABP_mov_phi}
\end{align}
with P\'eclet number $\Pe$ and  active reorientation strength $\OmA$, 
\begin{equation}
	\Pe = v_0/(r_H D_R) \ \ , \ \ \OmA = C_0/D_R ,
\end{equation}
the unit vector ${\bf e}_{\phi} = (- \sin \phi, \cos \phi)^T$, $r_c=|{\bf r} - {\bf r}_T|$, and the noise 
correlation functions $\langle \eta_{\kappa}(t) \eta_{\kappa}(t')\rangle = \delta(t-t')$, $\kappa \in \{r,\theta,R\}$. 
The  active reorientation of the pursuer toward a  target (Eq.~\eqref{eq:ABP_mov_phi}) is of the form of 
interactions applied in various other systems, e.g., the classical XY-model \cite{Kosterlitz1973}, Kuramoto 
oscillators \cite{Acebron2005}, and minimal cognitive models in swarming \cite{Barberis2016}.

We first consider the case where the pursuer moves much faster than the target,  corresponding to the limit 
of a stationary target which we place at the origin of the reference system, i.e., ${\bf r}_T =0$.   In the noise-free 
limit, $\Pe \gg 1$ and $\OmA \gg 1$, the stability analysis of Eqs.~\eqref{eq:ABP_mov_r} - \eqref{eq:ABP_mov_phi} 
yields the marginally stable fixed points (see Supplemental Material \cite{SI}) 
\begin{align}
r _0= \Pe/\OmA, \quad \beta_0 =  \theta_0 - \phi_0 - \pi =\pm \pi / 2 ,
\end{align}
which corresponds to circular trajectories with the radius $r_0 = \Pe/\OmA = r_H v_0/C_0$  (red circle in 
Fig.~\ref{fig:schematics}(b)). Surprisingly, the fixed point conditions neither  correspond to 
the solution where the pursuer is located at the target position, nor to a configuration where  the pursuer is 
oriented toward the target. Notably, the circular trajectory emerges by the specific form of self-propulsion, 
in which the pursuer cannot stop and reorient, but moves with a constant magnitude $v_0$ of the velocity --- as 
it is characteristic for ABPs. Numerically,  quasi-periodic orbits are obtained, which ``oscillate'' 
around the circle of radius $r_0$ (Fig.~\ref{fig:schematics}(b)), depending on the initial conditions when 
$r \neq r_0$---reminiscent of the planetary motion around the sun with perihelion rotation.  Thus, 
without noise, the pursuer can never reach  the target! Instead, noise is needed to kick the pursuer out of a 
quasi-periodic orbiting motion. 

\begin{figure*}
	\includegraphics[width=17.2cm]{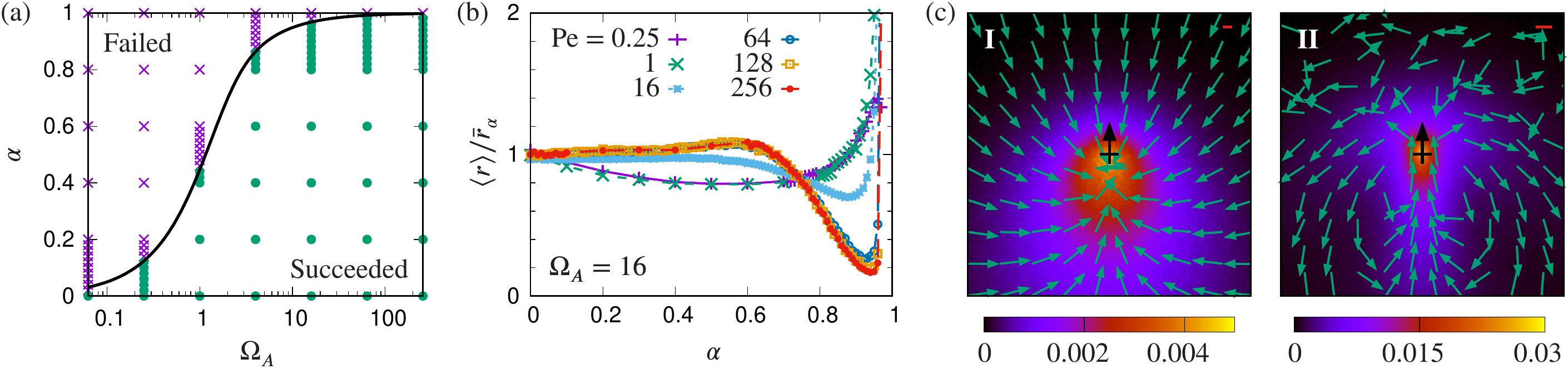}
	\caption{\label{fig:straight} 
		(a) Phase diagram of the pursuit of a linearly moving target. Symbols indicate failed ($\times$) and successful ($\CIRCLE$) pursuit. The P\'eclet number is  $\Pe = 16$ and the initial distance between 
		the target and the pursuer is $r_{\rm init} = 10^5$. The black line depicts  $\alpha_0$ of Eq.~\eqref{eq:cosbeta}.
		(b) Mean distance $\langle r \rangle$  between target and pursuer scaled by $\bar r_{\alpha}  = 2 \Pe/\OmA [1+\OmA/(\Pe^2(1-\alpha))]$ as a function of $\alpha$ for various P\'eclet numbers (symbols). The lines are guides for the eye. (c) Density distribution and 
		orientation of the propulsion directions for $\Pe = 0.25$ (I),  $\Pe =64$ (II), $\alpha = 0.6$, and $\OmA = 16$. 
		The target position is indicated by a plus (+) and the moving direction by an arrow.  The red scale bars 
		indicate the Stokes radius of the pursuers (cf. Supplemental Material, Movies S3 and S4 \cite{SI}).
		} 
\end{figure*}

The presence of noise changes the pursuer dynamics qualitatively, with two distinct behaviors depending on the 
P{\'e}clet number. For $r_0 \ll 1$, i.e., $\Pe \ll 1$, a pursuer preferentially aligns toward the target, 
and $\langle \cos \beta \rangle \lesssim 1$ for $\OmA \gg 1$, where $\beta= \theta- \phi  -\pi$ is the supplementary angle  to the angle $\theta-\phi$ between the vectors $\bf r$ and $\bf e$ (Supplemental Material, Movie S1~\cite{SI}). Then, the Fokker-Planck equation~\cite{Risken1989} for the radial 
distance yields the radial probability distribution function  $P(r) \sim e^{-2 r/\langle r \rangle}$ with the 
average radial distance $\langle r \rangle = 2/\Pe$ (Supplemental Material \cite{SI}).  In the opposite limit $\Pe \gg 1$ and $\OmA \gg 1$, we again find an exponential 
distribution function for $r$,  but now with the average  $\langle r \rangle = 2 r_0$. Here,  a pursuer traverses 
rosette-like trajectories as in the noise-free limit, however, now perturbed by noise with an essentially 
uniform distribution of the angle $\beta$ (Supplemental Material, Movie S2~\cite{SI}). The average radius of both dynamical regimes is well described by the 
expression    
\begin{align} \label{eq:rave}
\bar r  = 2 r_0 \left( 1+1/(r_0^2 \OmA)\right),
\end{align}
as shown in Fig.~\ref{fig:stationary}(a), which interpolates between the predicted limits. The minimum distance 
in terms of $\Pe$ and $\OmA$ follows for $\Pe/ \sqrt{\OmA} =1$.  This demonstrates that strong self-propulsion 
not necessarily enhances pursuit performance, but rather that magnitudes of propulsion and reorientation have to
work in unison. Here, noise plays an important role, because $\Pe/ \sqrt{\OmA}$ depends on $D_R$. 

To characterize the  influence of noise further, a mean first-passage time~\cite{Redner2001} is calculated as the average time of 
a pursuer starting within a circle of radius of $r_H$ centered at the target and returning for the first time 
again to the circle. Examples of the obtained probability distributions of return times are provided in the Supplemental Material \cite{SI}.  The mean first-passage time $\tau_{\mathrm{fp}}$ is then calculated as average of the return time starting at 
a shortest time, which is chosen as the time where the pursuer mean-square displacement exceeds the radius $r_H$.  
Figure~\ref{fig:stationary}(b) displays the mean first-passage time as a function of the P\'eclet number and 
various $\OmA$. Interestingly, the $\tau_{\mathrm{fp}}$ data for the various $\Pe$ and $\OmA$ all collapse onto 
an universal scaling curve when plotted as a function of the ratio $\Pe/\sqrt{\OmA}$ . 
The solid line in Fig.~\ref{fig:stationary}(b) presents a interpolation with the power-laws 
$\langle \tau_{\mathrm{fp}} \rangle \sim \Pe^{-9/5}$ for $\Pe/ \sqrt{\OmA} <1$ and 
$\langle \tau_{\mathrm{fp}} \rangle \sim \Pe^{4/5}$ for $\Pe/ \sqrt{\OmA} >1$. Consistent with the minimal mean 
pursuer-target distance for $\Pe/\sqrt{\OmA} =1$, the mean first-passage time is also minimal for the same value.  

For moving targets, the pursuer dynamics changes qualitatively. We examine first the  case, where the  target 
moves along a straight line $\mathbf{r}_T = (u_0\,t) \,\mathbf{e}_y$ with constant velocity $u_0$. The pursuer 
position is characterized in a co-moving reference frame of the target, i.e., ${\bf r}_T \equiv 0$ and  
${\bf r}  \equiv {\bf r}_c$. The fixed point of the noise-free equations of motion is then $\theta = - \pi/2$ 
and $\phi = \pi/2$, i.e., $\beta=0$ and the pursuer points directly toward the target (Supplemental Material \cite{SI}). The pursuer follows the target at a constant distance when the velocity ratio $\alpha =u_0/v_0$ is unity. Similarly, in the absence of noise, the pursuer follows the target as long as $\alpha <1$ along initial-condition dependent trajectories, comparable to a stationary target. In contrast, for $\alpha > 1$ the distance between target and pursuer diverges and pursuit fails, both with and without noise.

A ratio $\alpha < 1$ is particularly important in the presence of noise, because then a pursuer trajectory can 
never be as straight as the target trajectory, so that a speed $v_0> u_0$ is required to reduce
the distance to the target at large distance ($r \gg 1$). In the limit $r/r_0 \gg 1$, the dynamics of the iABP propulsion direction, $\phi$,  decouples from that of the iABP position, and we find 
\begin{align} \label{eq:cosbeta}
\langle \cos \beta \rangle =  I_1 (\OmA)/I_0 (\OmA) \equiv \alpha_0(\OmA) \ ,
\end{align}
where $I_0(\OmA)$ and $I_1(\OmA)$ are  modified Bessel functions of the first kind  (Supplemental Material, Sec.~S-II A \cite{SI}). The equation of 
motion for the radial distance leads then to the condition $\alpha < \alpha_0(\OmA)$ for a successful pursuit, which 
is confirmed by our simulations as shown in  Fig.~\ref{fig:straight}(a).  Here, the dynamical behavior for small
$\Pe \ll 1 $ is very similar to that of a stationary target, with an exponential distribution 
of the distance $r$, but with the mean value $\langle r \rangle = 2/[\Pe(1-\alpha)]$.  In particularly for 
$\OmA \gg 1$,  $\alpha_0 \lesssim 1$ and the pursuer is preferentially orientated toward the target (Fig.~\ref{fig:straight}(c)-I). 

In the limit of large $\Pe \gg 1$,  the pursuer overshoots the target, specifically for $\alpha \ll 1$ (Fig.~\ref{fig:straight}(c)-II). As for a 
stationary target, we find  an exponential distribution of the radial distance with $\langle r \rangle = 2 r_0$.  Overall,  the mean distance is  well described by the expression $\bar r_{\alpha}  = 2 \Pe/\OmA [1+\OmA/(\Pe^2(1-\alpha))]$ as long as $\alpha \lesssim 0.7$ (Fig.~\ref{fig:straight}(b)),  with the additional dependence on $\alpha$ compared
to Eq.~\eqref{eq:rave} (see also Supplemental Material, Fig.~S5(a)~\cite{SI}).
For $\alpha \gtrsim 0.7$ and $\Pe \gtrsim 10$, we find substantial 
deviations from $\bar r_\alpha$, and $\langle r \rangle$ assumes a minimum  in the vicinity of $\alpha_0$ before it diverges in the limit $\alpha \to 1$.  Configurations for $\alpha$ values in the vicinity of the minimum are similar to those  at the fixed point, $\alpha=1$, yet with $\alpha<1$ due to noise (Supplemental Material, Movie S5~\cite{SI}).

\begin{figure}[t]
\includegraphics[width=8.6cm]{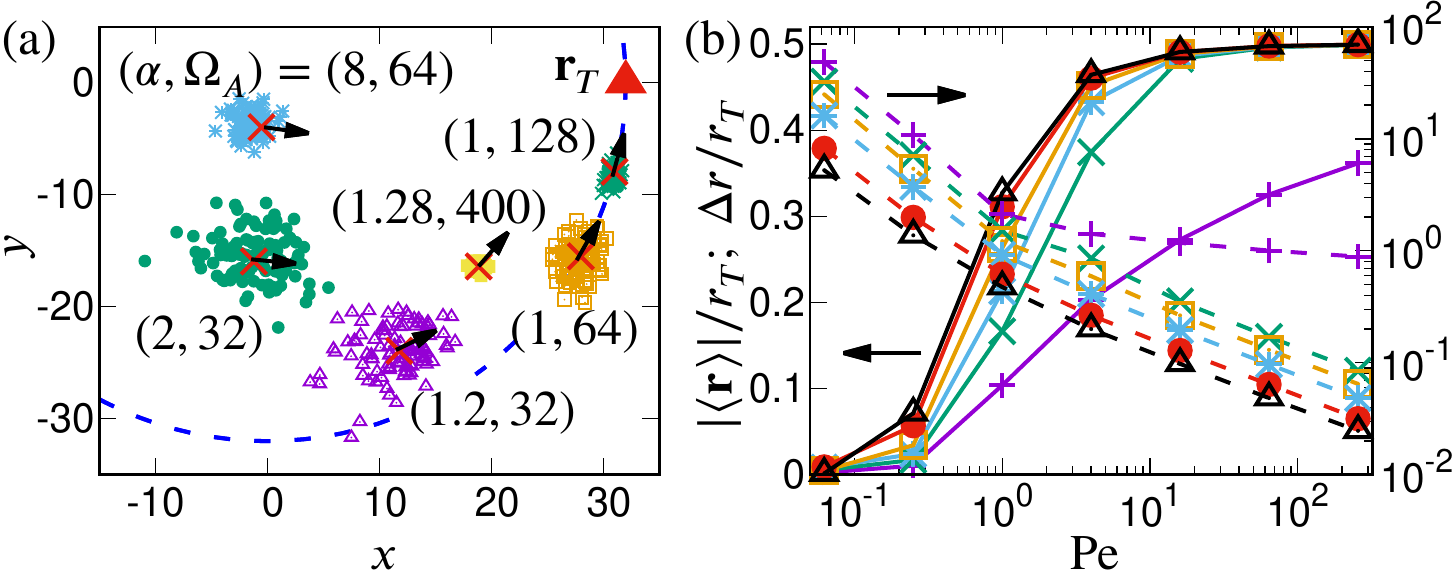}
\caption{ \label{fig:circular} 
 Target on circular trajectory. (a) Groups of pursuers moving on circles following the target located at ${\bf r}_T = (32,0)^T$ (red triangle) in a co-rotating reference frame with origin at $(0,0)$ for various pairs ($\alpha, \OmA$), $u_0/(r_H D_R) =512$, and $\OmT =16$. The dashed line illustrates the circle for $\alpha=1$. The black arrows indicate the average moving direction of a group, and the red crosses the theoretical prediction  Eq.~(S21)  (cf. Supplemental Material, Movies S6, S7, and S8 \cite{SI}).
  (b) Mean distance (left axis) and the root-mean square  fluctuations (right axis) as a function of the 
 P{\'e}clet number for various target radii: $r_T=$2 ($+$), 3 ($\times$), 4 ($\square$), 6 ($*$), 12 ($\CIRCLE$), and 20 ($\triangle$). Here, $r_0 = 1$  and $\alpha  = 2$.
 }
\end{figure}

Lastly, we explore the role of the shape of the target trajectory on pursuit by studying a target moving on 
a circular trajectory with angular velocity $\OmT$ and the radius $r_T$, i.e.,
$\mathbf{r}_T = r_T (\cos{(\OmT t)}, \sin{(\OmT t)})^T$. The noise-free equations of motion  yield a stable 
fixed point with the angle  $\beta =\pi/2$ and  characteristic pursuer radius $\tilde r = \Pe/\OmT$ 
(Supplemental Material \cite{SI}). Again, the dynamics depends on the ratio of the target and pursuer velocity 
$\alpha  = u_0/v_0 =  r_T \OmT /v_0 = r_T/\tilde r$, with the target velocity $u_0= \OmT r_T$. In the limit of 
$\alpha \ll 1$, the  pursuer radius is much larger than the target radius, and the equations of motion reduce again 
to those of a fixed target. 

For $\alpha > 1$, the target moves faster than the  pursuer. However, the distance $r_c$ does not diverge, 
because of the target's circular trajectory. Pursuers are able to keep track, but in a strikingly different 
manner from straight moving targets. Since $\alpha = r_T/ \tilde r > 1$, pursuers are located inside the 
circle of the target trajectory, where they move shorter distances than the target and, hence, are able to 
remain close to it. As the  fixed point is stable,  pursuers are localization in its vicinity  and move on circles with a  mean radius $\tilde{r}=r_T/\alpha$, only perturbed by noise (Supplemental Material, Sec.~S-1~C~\cite{SI}). This is confirmed by our simulations and, as depicted in Fig.~\ref{fig:circular}(a),  the fixed-point prediction describes the 
pursuit dynamics very well. The increasing influence of noise for $\Pe < 1$ causes deviations from the analytical prediction, in particular for small  $r_T$ (Supplemental Material, Fig. S1(b)).
The mean position is shifted toward the origin of the target circle as noise becomes more important, which allows pursuers to reduce the path length and to follow the target. Still, accumulation of slow pursuers inside 
the target circle is robustly observed in simulations. In addition, the fixed point analysis predicts a phase shift between the  instantaneous target angle $\OmT t$  and the pursuer angle $\theta (t)$, which depends on $\alpha$ and the ratio $\OmA/\OmT$. This is demonstrated in Fig.~\ref{fig:circular}(a), which shows a decreasing shift with increasing $\OmA$ at a given $\alpha$, consistent with the theoretical expression (Supplemental Material, Sec.~S-1~C~\cite{SI}).  

Noise also affects the distribution of particles in a group of non-interacting pursuers around their mean position, and leads to a radial repositioning of iABPs  with respect to  the radius $\tilde r$ during their circular motion. This is similar to repositioning of birds in a flock during a turn, and is a consequence the nearly constant pursuer speed, which implies a comparable length and curvature  of the traveled paths of the iABPs \cite{Hemelrijk2011}. As a consequence, a clockwise rotation of the pursuers' propulsion direction occurs in a reference frame rotating  counterclockwise with the frequency $\OmT$ (Supplemental Material, Fig. S1(d) and Movie S8~\cite{SI}).  

Figure \ref{fig:circular}(b) emphasizes the influence of noise on the mean radius $|\langle \bf r \rangle|$ 
of a group of (noninteracting) pursuers for various  target radii and P\'eclet numbers for $r_0=1$ and $\alpha =2$. Here, 
$|\langle \bf r \rangle|$ is the radius of the center-of-mass position $\langle \bf r \rangle$ of the pursuer 
group.  Due to noise, the mean radius is very small in the limit $\Pe \to 0$, but increase with  increasing 
$\Pe$ and saturates at the noise-free, large P\'eclet-number limit 
$\langle r \rangle = \tilde r = r_T/\alpha =1/2$ for the considered $\alpha$. The convergence toward the 
limit cycle of the noise-free case with increasing $\Pe$ is reflected in the  root-mean-square 
fluctuations $\Delta r$ with respect to  $|\langle \bf r \rangle|$, which exceed $|\langle \bf r \rangle|$ 
by far for $\Pe <1$ and  decrease with increasing $\Pe$.  

As our studies show, targets moving on circular trajectories can be employed for pattern formation and particle 
sorting according to the propulsion strength and reorientation capability. This can be seen as an 
``inverse herding'' effect, where a fast-moving circling sheep keeps a herd of slower shepherd dogs together, 
which are all chasing the sheep.

In summary, we have considered noisy pursuit dynamics of iABPs, for various types of target trajectories.
We have shown that noise plays a dual role in the successful approach of the target. 
On the one hand, it is required to kick the pursuers out of regular, quasi-period orbits around 
the target, and thus facilitates  target approach. On the other hand, noisy trajectories are longer 
than straight trajectories, and, hence, slow down the pursuit,  requiring higher pursuer velocities.
Finally, we have demonstrated that the geometry of target trajectories can be employed for 
pattern formation and sorting of active agents according to their motility.


%

\end{document}